\documentstyle[11pt,hrd_pasp,twoside,epsf]{article}
\def\edcomment#1{\iffalse\marginpar{\raggedright\sl#1\/}\else\relax\fi}
\reversemarginpar

\begin{document}
\title{CMD's, RR Lyrae's, clump stars and reddening in the LMC}
 \author{Gisella Clementini, Angela Bragaglia and Marcella Maio}
\affil{Osservatorio Astronomico di Bologna, via Ranzani 1, 40127 Bologna, Italy}
\author{Eugenio Carretta and Raffaele Gratton}
\affil{Osservatorio Astronomico di Padova, Vicolo dell'Osservatorio 5, 
35112 Padova, Italy}
 \author{Luca Di Fabrizio}
 \affil{Telescopio Nazionale Galileo, 38700 Santa Cruz de La Palma, Spain}

\begin{abstract}

We present new {\it B, V, I} photometry in the standard Johnson-Cousins 
photometric system of two regions near the bar of the Large Magellanic
Cloud (LMC), one of which is very close to LMC
field 
used for the "Coimbra test on the synthetic CMD 
techniques".
The two areas contain 128 RR Lyrae variables and more than 
8000 clump stars.
Once combined with Clementini et al. 2001
(astro-ph/0007471) photometry the new data-set allows us to get 
(i) accurate CMD's down to $V\sim$23 mag, 
(ii) full coverage of the {\it B} and {\it V} light variation for more than 
80\% of the RR 
Lyrae's, (iii) a very precise estimate of the LMC RR Lyrae's average apparent 
luminosity, (iv) an accurate estimate of the {\it I} luminosity of the 
LMC clump stars, and (v) an independent evaluation of the reddening in these 
regions of the LMC using the 
pulsational properties of the RR Lyrae's in our sample.
A detailed comparison is made between our, OGLE II and MACHO photometries.
We discuss the impact of our new results on the distance to the LMC and 
on the {\it short} and {\it long} distance scale controversy.

\end{abstract}

\section{Introduction}
Despite its vicinity and notwithstanding the enormous 
observational effort devoted to its study, many of the LMC properties 
(distance, reddening, star formation history, etc.) are not yet sufficiently 
well-known. Among them, the distance is by far the most 
controversial issue with a 0.2-0.3 mag difference between {\it short}
and {\it long} scales provided by different distance indicators 
found to coexist in the 
LMC (e.g. Cepheids, RR Lyrae, clump stars, etc.).
In the {\it short} distance scale, which is the scale favoured by the 
statistical parallaxes of field RR Lyrae variables and 
the Clump method, the 
distance modulus to the
LMC  is about 18.24 mag (Udalski 2000), the age of the galactic Globular
clusters (GGCs) is about 16 Gyr, and the absolute luminosity of the 
RR Lyrae is about 0.7 mag.
In the {\it long} scale, which is the scale favoured by Cepheids and 
Main Sequence Subdwarf Fitting of GGCs, the distance modulus of the 
LMC is about 18.5 mag, the age of the GGCs is 13 Gyr, and the absolute 
magnitude of the RR Lyrae is about 0.5 mag (Carretta et al. 2000, and
references therein).

Thanks to their pulsational properties combined with the 
almost constant absolute magnitude RR Lyrae variables have long been 
recognized to be excellent tracers of old stellar populations 
as well as primary distance indicators for the Local Group 
galaxies. About 8000 field RR Lyrae have been discovered in the LMC by the 
MACHO microlensing experiment (Alcock et al. 1996). 
We have observed two regions of the LMC bar 
which according to Alcock et al. (1996) were expected to host 
many RR Lyrae as well as a large number of 
clump stars. 
These data allowed us to 
study the photometric and
pulsational properties of the RR Lyrae which fall in the 2 
areas in great detail, to get a very precise estimate of   
the apparent luminosity of the Horizontal Branch (HB) and to 
measure the average 
metallicity of the LMC old stellar 
component as defined by its RR Lyrae stars. Finally we have also 
obtained an independent evaluation of the reddening 
in these fields of the LMC bar.
The comparison between average luminosities of RR Lyrae and  
clump stars in the same area also allowed us a direct confrontation 
of the related distance scales. Our results directly bear upon the "Coimbra experiment",
since provide independent informations on 
the distance modulus, the
metallicity of the old stellar component and the reddening value, 
in the region of the bar of the LMC simulated by the different 
CMD techniques.

\section{Observations and data reductions}

Time series {\it B,V,} and {\it I} exposures of two 13 $\times$ 13 
arcmin$^2$ fields 
(namely, Field A and B) close 
to the bar of the LMC and contained in Field \#6 and \#13 of the MACHO 
microlensing experiment (Alcock et al. 1996) have been obtained with the
1.5 Danish telescope at ESO, La Silla, in two observing runs 
respectively in January 1999 and 2001. Field A also has a 40\% overlap 
with OGLE II field LMC\_SC21 (Udalski et al. 2000) so that a star by star 
comparison is possible, for the first time, between these 
two photometries. 
The full data-set consists of 72, 41 and 15 frames in {\it V, B} and {\it I}
respectively.
152 variables were identified in the two fields of which 128 
are RR Lyrae's. Full coverage of the light curve has been obtained 
for more than 80\% of them.
We have also obtained low resolution spectra and measured the $\Delta$S metal 
abundances (Preston 1959) for 6 of 
the 10 {\it RRd} falling in the sample, using EFOSC2 at the 3.6 m telescope. 
Data reduction of the photometric frames was performed using 
DoPHOT (Schechter, Mateo \& Saha 1993) for the 1999 observations, and 
the DAOPHOT and ALLFRAME
reduction packages (Stetson 1987, 1994) for the 2001 data.

Panel {\it a} of Figure 1 shows the  {\it V vs (B$-$V)} diagram of 
Field A from data taken in 1999, the remaining 3 panels show the {\it V vs (B$-$V)},
{\it V vs (V$-$I)}, and {\it I vs (V$-$I)} diagrams for the 2001 data.
Beside the higher sensitivity in {\it V} and {\it B} of the CCD used in 2001
(an EEV chip in 2001 and a Loral-Lesser CCD in 1999)
and the better seeing and photometric conditions during the observations, a  
further improvement upon the 1999 photometry was achieved thanks to the use 
of the reduction packages
DAOPHOT and ALLFRAME. The new HR diagrams reach about one magnitude fainter 
(V$_{limit}\sim$23) with about 3 times as stars than in the 1999 data-set.
{\it V, B}, and {\it $\Delta$i} light curves for some of the variables in our sample are 
shown in Figure 2. Perfect agreement is found between light curve data 
points corresponding to the two separate runs.

\begin{figure}
\plotfiddle{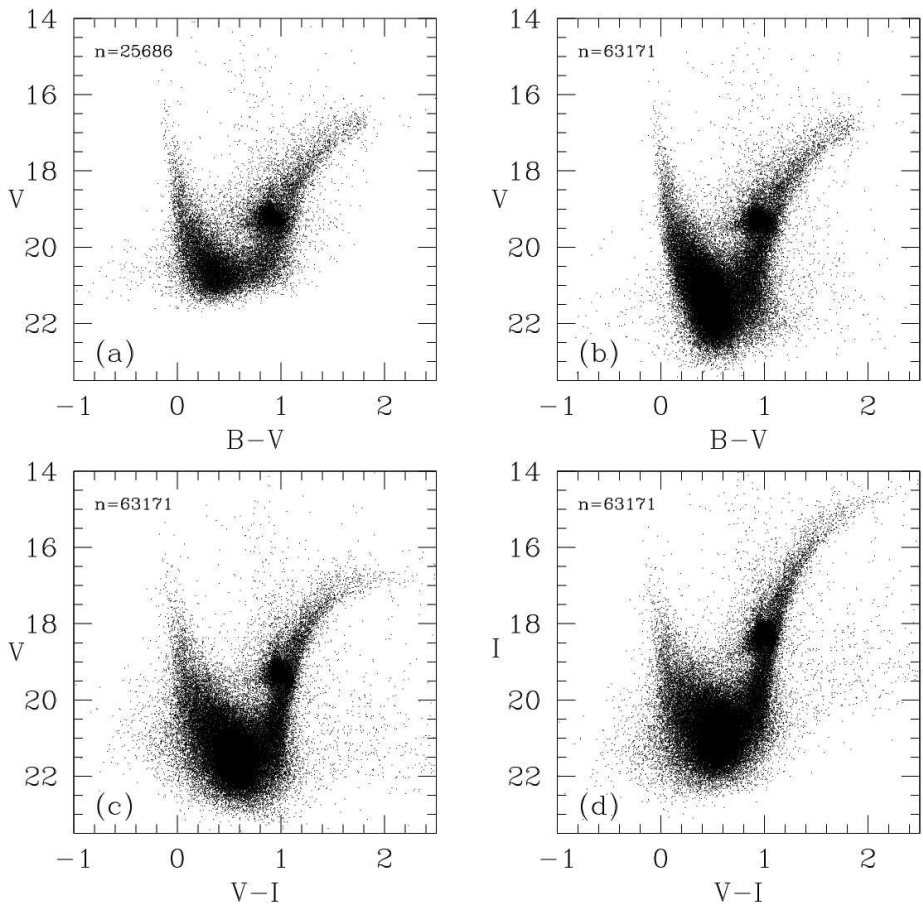}{12cm}{0}{65}{65}{-210}{-15}
\caption{HR diagram of LMC Field A, from data in 1999 (panel {\it a})
and in 2001 (panels {\it b, c, d}), respectively.}
\end{figure}

According to the period distribution of the RR Lyrae's in our sample,
the LMC has properties intermediate between the Oosterhoff types I and II
(Oosterhoff 1939).
There is a  small difference in the average luminosities of the RR Lyrae
in the two fields which suggests that a 0.02 mag differential reddening may exist 
between the two areas. 

The reddening is known to be patchy and vary
from one region to the other of the LMC. A large number of different 
estimates exist in the literature with values in the range 
from 0.03 to 0.22 mag, depending on the region of the LMC and on 
the reddening indicator adopted.
Since reddening is a major ingredient in any distance derivation 
procedure, it is particularly important 
to estimate the reddening
in the same region of the distance indicators used to measure the distance to the 
LMC. 
Reddening in our two 
fields was derived from the comparison of the colours of the edges of the 
instability strip defined by 
the RR Lyrae in our sample, with those of the Galactic globular cluster 
M3 ([Fe/H]=$-$1.5, E(B$-$V)=0.00 mag) 
based on Corwin \& Carney (2001) photometry.
We derive values of 0.09 and 0.07 mag for Field A and B, respectively, and 
an upper limit average value of 0.10 mag. Our estimate is 0.044 mag smaller 
than derived in the same area 
by Udalski et al. (1999) using the {\it I} luminosity of the
clump stars.
The average properties, the metallicity (Bragaglia et al. 2001) 
and the reddening inferred from the RR Lyrae in our sample are summarized in Table 1.

\begin{table}
\caption{Average properties of the LMC RR Lyrae in our sample}
\vspace*{3mm}
\begin{tabular}{lccccc}
\tableline
$<$P$_{ab}>$& Oo-type&V(RR)$_{\rm Field A}$&V(RR)$_{\rm Field B}$&
[Fe/H]&E(B$-$V)\\[2pt] \hline\\[-10pt]
0.573&OoI/OoII&19.369$\pm$0.023&19.314$\pm$0.025&$-1.5 \pm 0.2$&0.10\\
\tableline
\end{tabular}
\end{table}

\begin{figure}
\plotfiddle{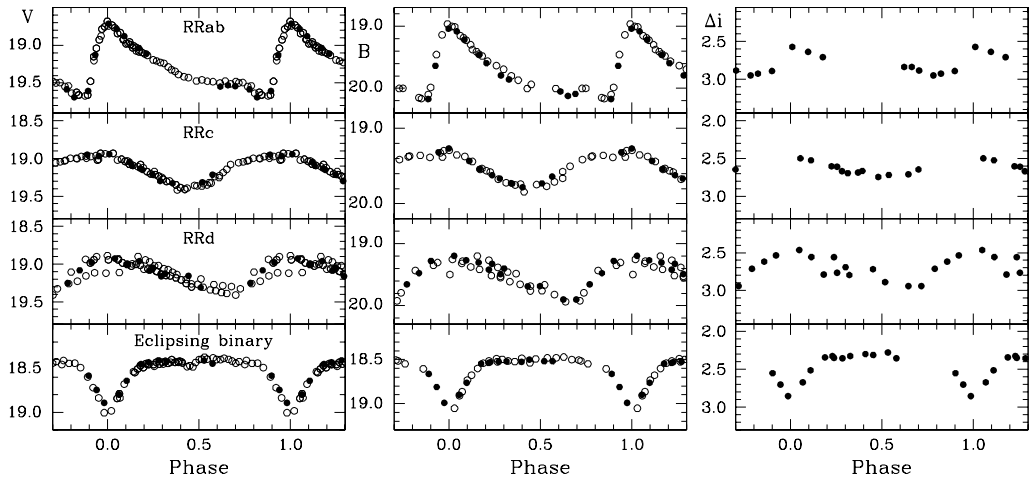}{7cm}{0}{70}{70}{-210}{-150}
\caption{Some 
examples of our light curves. Open circles are from Clementini et al. (2001), 
filled dots are the new data points. Only differential light curves in the 
instrumental magnitude system are shown for the {\it I} band, since 
the calibration of the {\it I} data is still in progress.}
\end{figure}
The comparison between the average luminosity of the RR Lyrae in our sample 
and literature values are summarized in Table 2.

\begin{table}
\scriptsize
\caption{Comparison with the $<$V(RR)$>_0$ values in the literature
 transformed to a common value of the metallicity: [Fe/H]=$-$1.5}
\vspace*{3mm}
\begin{tabular}{crclcccc}
\tableline
$<$V(RR)$>_0$&N$_{\rm stars}$&\multicolumn{4}{c}{Error}&Reddening&Reference\\
&&(1)&~~(2)&(3)&(total)&&\\
\tableline
19.03&~93&0.02&~0.03&0.06&0.07&0.10&Present work\\
19.03&182&0.04&0.025&0.05&0.07&0.09&Walker (1992)\\
19.095&500& --   &0.073&0.06&0.10&0.1&MACHO (4)\\
18.92&71&0.04&0.015&0.08&0.09&
0.144&OGLE II LMC\_SC21 (5-6)\\
18.91&--&0.04&0.015&0.08&0.09&0.143&OGLE II (7)\\
\tableline
\end{tabular}
(1)= standard deviation of the average, (2)=photometric zero point, 
(3)= absorption contribution
(4)= Alcock et al. (1996), (5) Udalski et al. (1999), (6) Udalski et al. (2000),
(7)= Udalski (2000)
\normalsize
\end{table}

Our dereddened average luminosity of the RR Lyrae is in very good agreement 
with Walker's (1992) value for the 
LMC cluster RR Lyrae's, and is 0.06-0.07 mag brighter than MACHO's.
This difference may be due to the 
unconventional filter bands of the MACHO photometry.
OGLE II's  $<$V(RR)$>_0$ is about 0.10 mag brighter than our 
estimate due to higher reddening
value adopted by these authors. 
A star by star comparison is also possible with OGLE II based 
on the more than 5000 clump stars we have in common. 
Particularly important is the comparison of the 
average luminosities of RR Lyrae and Clump stars in the 
{\it I} band, 
since the {\it I} luminosity of the Clump is 
the method which provides the shortest distance to the
LMC ($\mu _{\rm LMC}$(Clump)=18.24 mag, Udalski 2000).  
Our average {\it I} luminosity for the clump stars is 18.28 mag,  
to compare with Udalski's (2000) 18.25 mag. 

However, it should be 
reminded that the Clump of 
a composite population, as the one in the LMC, has a complex structure
resulting from the superposition of stellar populations with different 
masses and ages.
Strong evolutionary and age effects are then present and should be properly 
accounted for when the Clump is used as a distance indicator.
On the other hand, the contribution to the Clump by the horizontal branch (HB) of the {\it old stellar population}
in the LMC is very well defined by its RR Lyrae's. The average luminosity of the RR Lyrae 
very clearly corresponds to the lowest envelope of the Clump. 
This is shown in Figure 3, that displays the enlargement of the HB 
and clump regions in the HR diagrams of our two fields.
In each panel of the figure the solid lines show the luminosity level of the 
HB of the old stellar component inferred from the average luminosity of 
the RR Lyrae stars.
In particular, if we use the RR Lyrae to trace 
the average {\it I} luminosity of the Clump of the old stellar component 
we derive a distance modulus for the LMC about 0.15 mag
larger than obtained by Udalski (2000), and the adoption of our reddening value
(E(B$-$V)=0.10 mag) would contribute further to lengthen by 
0.14 mag the distance modulus of the LMC.

Finally, if we  assume for the absolute magnitude of the RR Lyrae either the 
HB luminosity derived from the 
new and still unpublished Main Sequence Fitting distances of NGC6397 and NGC6752
 by Gratton and collaborators, based 
on their ESO-VLT Large Program on GGCs, or, alternatively,
values from the revised Baade-Wesselink 
analysis of Cacciari et al. (2000), 
and combine them with our $<$V(RR)$>_0$
we obtain a distance modulus 
for the LMC of 18.47 mag, in better agreement with the "long" distance scale.
\begin{figure}
\plotfiddle{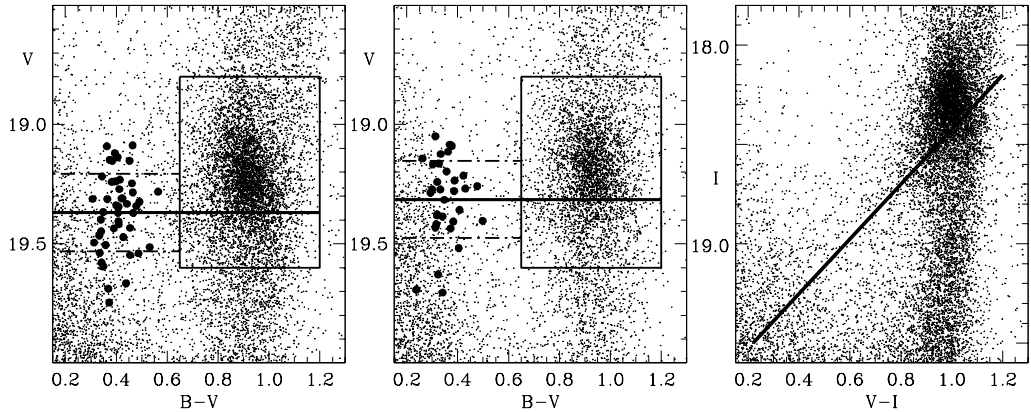}{5.5cm}{0}{70}{70}{-210}{-160}
\caption{Enlargement of the HB and Clump 
regions in the {\it V vs B$-$V} HR diagram of Field A (left panel) and B (central panel), respectively, from data
taken in 1999. In the right panel the {\it I vs V$-$I} HR diagram of Field A from the 2001 data.
Filled circles mark the RR Lyrae stars which are plotted according to their intensity average 
magnitudes and colors. In each panel solid lines show the luminosity level of the 
HB of the old stellar component inferred from the average luminosity of 
the RR Lyrae stars. The dot-dashed lines correspond to the 1 $\sigma$ deviations
from the average.}
\end{figure}

\end{document}